# *Getting Started with PATSTAT Register*

Gaétan de Rassenfosse [*], Martin Kracker [#] and Gianluca Tarasconi [†]

[*] Ecole polytechnique fédérale de Lausanne, College of Management of Technology, Chair of Innovation and IP Policy. Station 5, 1015 Lausanne, Switzerland. gaetan.derassenfosse@epfl.ch

[#] European Patent Office, Electronic Publication and Dissemination. Rennweg 12, 1030 Vienna, Austria. mkracker@epo.org

[†] Invernizzi Center for Research in Innovation, Organization and Strategy (ICRIOS), Università Bocconi. Via Sarfatti 25, 20136 Milan, Italy.

**Abstract**

This paper provides a technical introduction to the PATSTAT Register database, which contains bibliographical, procedural and legal status data on patent applications handled by the European Patent Office. It presents eight MySQL queries that cover some of the most relevant aspects of the database for research purposes. It targets academic researchers and practitioners who are familiar with the PATSTAT database and the MySQL language.

# 1. Introduction

The PATSTAT Register is still a relatively unknown patent database but it is a real gem in terms of the richness of the data that it contains. In contrast to the core PATSTAT database, which contains worldwide data, the PATSTAT Register contains only information about patent applications at the European Patent Office (EPO), but at a considerably deeper level. It contains bibliographic, procedural, and legal status data on published European and Euro-PCT patent applications. We direct readers with limited knowledge of the PATSTAT database to de Rassenfosse, Dernis and Boedt (2014).

We assume that the reader is familiar with the MySQL query language and that both the "core" PATSTAT database and the PATSTAT Register database are up and running.[1] We have used the 2015 Spring Edition of the core PATSTAT database and the 2016 Autumn Edition of the PATSTAT Register database. One query also relies on the INPADOC legal status table (called TLS221_INPADOC_PRS, but that will be replaced by TLS231_INPADOC_LEGAL_EVENT in future versions of PATSTAT). The PATSTAT Register database mainly covers information related to the application procedure of EP applications until the grant of the patent. By contrast, the INPADOC legal status table primarily covers legal status data of granted patents from more than 40 patent authorities worldwide, including the EPO.

There are only a handful of research projects that have exploited European Patent Register data. Harhoff and Wagner (2009) provide an early use of the Register data. They rely on the online EP Register application to study the determinants of the duration of patent examination at the EPO. van Zeebroeck and van Pottelsberghe (2011) rely on data obtained directly from the EPO to study the determinants of patent value, including information on oppositions. Gäßler and Harhoff (2016) use EP Register data together with data from the German Patent Office to build a dataset of change in ownership for patents with legal validity in Germany. Bösenberg and Egger (2016) study the extent to which R&D tax incentives affect the trading of patents.

The paper presents eight queries, each covering interesting aspects of PATSTAT Register for research purposes. Although the paper does not cover all the tables and attributes available in the database, it does provide a sense of what the data look like. (For a complete technical description of the database, please consult the data catalogue available on the EPO website). We hope that it will help researchers to engage in new research directions.

Figure 1 provides a broad overview of the PATSTAT Register. The patent application is the central entity of the database. The attribute ID is the identifier for applications, and consequently it is part of the primary key of most tables. Similarly to PATSTAT's attribute APPLN_ID, the Register ID is stable between versions of the database. In the remainder of this document, we use capital letters in the main text to denote data tables and attributes.

---

[1] To discover these databases, the EPO freely distributes a dataset of patents related to wind turbine technologies. It is available at http://www.epo.org/searching-for-patents/business/patstat.html#tab4. The European Patent Register, an online application to look up Register data for single applications is freely available on the EPO website (try, *e.g.*, https://register.epo.org/application?number=EP07075884).

**Figure 1.** Overview of the PATSTAT Register

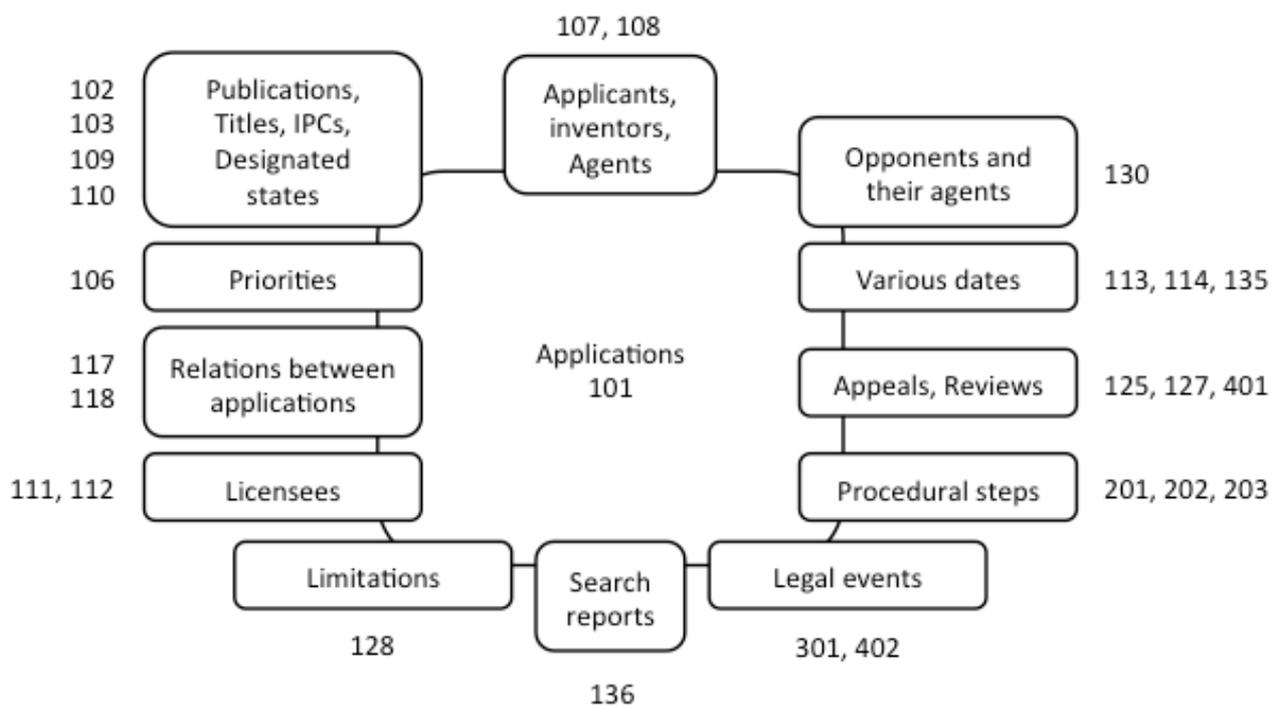

Source: Adapted from EPO promotion document.

## *2. Eight queries explained*

All the queries are executed on a sample of patent applications filed at the EPO between the years 2000 and 2010 and that were in the field of wind motors. The reasons for focusing on a specific sample are to increase comparability of the results presented in this paper with future releases of the database, as well as to minimize the execution time of queries.

We start by creating the table of wind motor patents using the PATSTAT database, as illustrated in Query 0. In brief, the query creates a table called WIND_PATENTS that lists the identifiers (attribute APPLN_ID) of patent applications filed at the EPO, including PCT applications, that have at least one of their IPC codes starting with F03D (denoting wind motors). There are 4375 applications in our version of the database. For pedagogical reasons we use the function YEAR() to extract the year information from attribute APPLN_FILING_DATE. An alternative option involves using directly attribute APPLN_FILING_YEAR. This practice is better from a technical standpoint as the year column can be indexed, which speeds up queries—by contrast, one cannot index a function of a column.

**Query 0**

```sql
CREATE TABLE wind_patents AS
SELECT
   DISTINCT t1.appln_id
FROM
   tls201_appln t1
      INNER JOIN
   tls209_appln_ipc t9 ON t1.appln_id = t9.appln_id
WHERE
   t1.appln_auth = 'EP'
      AND (t1.appln_kind = 'A' OR t1.appln_kind = 'W')
      AND year(t1.appln_filing_date) BETWEEN 2000 AND 2010
      AND t9.ipc_class_symbol LIKE 'F03D%'
```

*2.1 Linking PATSTAT Register to PATSTAT core tables*

It is straightforward to link records from PATSTAT Register to PATSTAT core tables. Query 1 shows a simple way to count the number of references listed in patent applications (sometimes referred to as backward citations). The citation count is quite basic and only serves the purpose of illustrating how the linking works.

The starting point of Query 1 is table REG101_APPLN. It contains basic data about applications such as the filing date, the patent office (which is always the EPO), the application number, and the status of the application (*e.g.*, patent revoked, application withdrawn, examination in progress). The query links records in REG101_APPLN to TLS211_PAT_PUBLN (via table WIND_PATENTS) using directly the PATSTAT identifier (APPLN_ID). In all the queries we will show both identifiers, in order to allow further elaborations either with PATSTAT core or with PATSTAT Register tables. However, it is important to note that the information is redundant, in the sense that there is a one-to-one correspondence between the PATSTAT Register identifier and the PATSTAT core identifier for patent applications that are in both databases.[2]

Note that it is good practice in MySQL (and required in other languages such as MS SQL) to include both the PATSTAT Register and the PATSTAT core identifiers in the GROUP BY clause because they both appear in the SELECT clause. Indeed, the database management system doesn't know that the information is redundant.

---

[2] About 30 per cent of all applications in the REG101_APPLN table are international applications that have not entered the EP regional phase. Since no EP publication exists in the PATSTAT core tables for these applications, the attribute REG101_APPLN.APPLN_ID will have value 0; thus, the PATSTAT Register database contains more EP applications than the PATSTAT database.

```sql
                                                                           Query 1
SELECT
   r101.id, r101.appln_id,
   count(DISTINCT t212.cited_pat_publn_id) AS n_cit
FROM
   tls212_citation t212
      INNER JOIN
   tls211_pat_publn t211 ON t211.pat_publn_id = t212.pat_publn_id
      INNER JOIN
   wind_patents wp ON wp.appln_id = t211.appln_id
      INNER JOIN
   reg101_appln r101 ON r101.appln_id = wp.appln_id
WHERE
   t212.pat_citn_seq_nr > 0
GROUP BY r101.id, r101.appln_id
ORDER BY n_cit DESC;
```

Table 1 reports the first five records. For example, ID 815670, corresponding to APPLN_ID 56608002 in the PATSTAT core database, cites a set of patent publications, which, together, make 86 references to prior patent literature.

**Table 1. First five records of output of Query 1**

| id | appln_id | n_cit |
|---|---|---|
| 8156970 | 56608002 | 86 |
| 10380117 | 328374009 | 34 |
| 5107140 | 16211141 | 29 |
| 9380102 | 57718882 | 20 |
| 13002613 | 406613665 | 19 |

*2.2 Assessing the geographic coverage of licenses*

PATSTAT Register contains some information on licensing. Patent licensing is an important channel through which new technologies diffuse (Shapiro 1985) and theoretical economists have studied questions such as the welfare implications of patent licensing under different industry structures. Empirical studies are less frequent in the literature, although certainly not absent from it (*e.g.*, Shane 2002, Gans, Hsu and Stern 2008). Scholars should keep in mind that not all technology-licensing transactions involve a patent. The holding of a patent facilitates licensing deals by protecting buyers against the expropriation of the idea but a technology can be licensed even if it is not protected by a patent. de Rassenfosse, Palangkaraya and Webster (2016) estimate that about 20% of technology transaction negotiations in Australia do not involve a patent.

PATSTAT Register contains the list of member states of the European Patent Convention (EPC) where a patent licensing has been applied. These data are contained in table REG112_LICENSEE_STATES, along with the year and bulletin number in which the agreement was announced. The license information is disclosed voluntarily to the EPO, such that the data are very sparse. However, it is a much sought-after information for research purposes.

Query 2 is quite simple. Table REG101_APPLN is the starting point for the query and allows getting the attributes ID and APPLN_ID. The table is then linked to REG112_LICENSEE_STATES using the ID attribute. Note the use of the DISTINCT clause in the COUNT instruction. Since it is possible to grant a license to more than one licensee, the DISTINCT clause avoids double counting of countries.

**Query 2**
```sql
SELECT
   r101.id, r101.appln_id,
   count(DISTINCT r112.licensee_country) AS nb_lic_ctry
FROM
   wind_patents wp
      INNER JOIN
   reg101_appln r101 ON wp.appln_id = r101.appln_id
      INNER JOIN
   reg112_licensee_states r112 ON r101.id = r112.id
GROUP BY
   r101.id, r101.appln_id
ORDER BY nb_lic_ctry DESC;
```

Table 2 reports the first five records of Query 2. Application ID 10788117 is associated with licensing agreements that cover 37 EPC member states, whereas licensing agreements for application ID 10742603 covers 36 EPC member states. Had we omitted the DISTINCT clause, the COUNT() would have returned 62 member states for application ID 10742603.[3]

**Table 2. First five records of output of Query 2**

| id | appln_id | nb_lic_ctry |
|---|---|---|
| 10788117 | 329924500 | 37 |
| 10742603 | 320770528 | 36 |
| 9737113 | 273166015 | 35 |
| 9785200 | 274118159 | 35 |
| 9009058 | 58007413 | 35 |

Finally, note that there are three main license types, which are reported in attribute REG111_LICENSEE.TYPE_LICENSE. The entry "EXC" corresponds to an exclusive license, the entry "NEX" corresponds to a non-exclusive license, and the entry "RIR" stands for "right in rem"—such rights may correspond, *e.g.*, to a claim for money damages (seizure, security agreement, mortgage). There are about 10,000 applications with an entry in table REG111_LICENSEE.

*2.3 Change in applicants and patent transfer*

The transfer of patents is another dimension of markets for technology. Serrano (2010) was among the first to present large-scale quantitative evidence on patent transfer. Using reassignment data from the USPTO, he finds that 13.5% of all granted U.S. patents are traded at least once over their lifecycle. Getting accurate information on patent transfer for Europe is particularly challenging, as there exists no centrally managed register. Gäßler and Harhoff (2016) focus on the specific case of patents with legal validity in Germany. Combining data from the German Federal Patent and Trademark Office and the EPO, they report that between 7 and 8% of the patents in their sample were transferred. Ciaramella, Martínez and Ménière (2016) tackle the issue of tracking transfers at the European level. They introduce the concept of "operations", which refers to parallel inscriptions of the same transfer across different registers (*e.g.*, in the EPO and the French registers). They then build an indicator of "transactions", which may involve several patents transferred at the same time between the same actors. Transfer data are important to study questions related, for example, to gains from technology trade and the division of innovation labor (*e.g.*, Arora and Gambardella 1994).

---
[3] In fact, table REG112_LICENSEE_STATE lists only licensee states if the attribute REG111_LICENSEE.DESIGNATION is "as-indicated" (meaning: see as indicated by table REG112_LICENSEE_STATES). If attribute REG111_LICENSEE.DESIGNATION is "all", then there will be no entry in REG112_LICENSEE_STATE, but the license covers all countries that were members of the EPC at that point of time.

It is possible to track changes of parties over time using table REG107_PARTIES, which contains data on applicants, inventors and legal representatives. The types of parties are distinguished by the attribute TYPE, and value "A" stands for "applicants". Two attributes are important in this respect. Attribute SEQ_NR numbers the applicants, if there are multiple applicants at the same time (similarly for inventors and agents). Attribute SET_SEQ_NR numbers the set of applicants that change over time. For every change of applicant there is a corresponding new set, *i.e.*, a new value for the attribute SET_SEQ_NR. For instance, ID 89117921 (APPLN_ID 16766302) initially had two applicants (SET_SEQ_NR = 1 and SEQ_NR = 1 and 2). At the first change, (SET_SEQ_NR = 2) there were still two applicants. Later (SET_SQ_NR = 3), there were three applicants. At the last change (SET_SQ_NR = 5 and IS_LATEST = Y) there were still three applicants.[4]

Query 3 simply identifies the number of changes in applicant information by identifying the maximum field of attribute REG107_PARTIES.SET_SEQ_NR associated with TYPE 'A'.

**Query 3**
```
SELECT
   r101.id, r101.appln_id, max(r107.set_seq_nr) AS nb_changes
FROM
   wind_patents wp
      INNER JOIN
   reg101_appln r101 ON wp.appln_id = r101.appln_id
      INNER JOIN
   reg107_parties r107 ON r101.id = r107.id
WHERE
   r107.type = 'A'
GROUP BY
   r101.id, r101.appln_id
ORDER BY
   nb_changes DESC, r101.id ASC;
```

The results are reported in Table 3. Application ID 3732247 has had six changes in applicant information before the patent was granted. The original applicant was "NEG Micon A/S" and the patent applicant was last changed on May 1st, 2015 to "Vestas Wind Systems A/S" (an online search suggests that the two companies merged in 2004). The last change in our version of the database corresponds to a change in address, from "Hedeager 44, 8200 Aarhus" to "Hedeager 42, 8200 Aarhus" (not reported).

---

[4] Note that applicants within the same cohort (= SET_SEQ_NR) may differ by their designation—that is, ownership is geographically defined. For example, the German subsidiary of a multinational may hold the DE/AT/CH-countries, while the holding remains the owner for the rest of the countries.

**Table 3. First five records of output of Query 3**

| id | appln_id | nb_changes |
|---|---|---|
| 3732247 | 16049513 | 6 |
| 4803066 | 16175353 | 6 |
| 6842801 | 56198546 | 6 |
| 106602 | 15711981 | 5 |
| 2025882 | 15918821 | 5 |

In theory, change in applicant information reveals that a patent has been transferred. In practice, not all changes are communicated to the EPO and not all communicated changes correspond to a genuine transaction. Thus, the data must be interpreted with caution. There are several (imperfect) approaches for identifying genuine changes in applicant. One approach involves looking at whether there was a change in field REG107_PARTIES.CUSTOMER_ID. However, parties have not been harmonized nor disambiguated such that they may have several customer identifiers.[5] Another approach involves looking at specific events in table REG301_EVENT_DATE. In particular, the four EVENT_CODES that end with "APPR" correspond to changes in applicant. However, again, caution is needed since an address change may be reported under a "Change – applicant" event and not a "Change – name/address" event.

*2.4 Measuring the time between filing date and examination date*

The time elapsed between the filing date and the examination date provides insights into the inner working of the patent office and reveals elements of firms' patenting strategy. For example, it can be used as a performance indicator for the patent office (Mitra-Kahn *et al.* 2013). But because applicants at the EPO have to ask for the request for examination (more precisely, pay the examination fees, which triggers examination), it also reveals elements of patenting strategy or the market potential of the invention.[6] For instance, Palangkaraya, Jensen and Webster (2008) provide evidence that applicants create investment uncertainty for their competitors by delaying decisions to request patent examination. As a matter of fact, there is considerable heterogeneity in the duration of patent examination at the EPO (Harhoff and Wagner 2009), and applicants have incentives to strategically delay or accelerate the prosecution process (de Rassenfosse and Zaby 2016).

Query 4 uses table REG301_EVENT_DATA that stores the full list of events that occurred to an application up to the end of the opposition period. After that point in time, the information about legal events is stored in the INPADOC legal status table (table TLS221_INPADOC_PRS or TLS231_INPADOC_LEGAL_EVENT). The attribute EVENT_CODE allows to identify the types of events. Code number '0009185' corresponds to the first examination report. (See table REG402_EVENT_TEXT for the descriptions of event codes used in REG301_EVENT_DATA).

---

[5] Contrary to PATSTAT TLS906_PERSON table, the CUSTOMER_ID attribute in PATSTAT Register does not rely on bibliographic information from the patent publications. It updates information even if the change does not trigger a new publication (which is only the case if there are amendments to the patent text).

[6] Applicants have a maximum of six months after the date on which the European Patent Bulletin mentions the publication of the European search report to pay the examination fees. After that time lag, the application is deemed withdrawn.

```
                                                                            Query 4
SELECT
   r101.id, r101.appln_id, r101.appln_filing_date,
   r301.event_date AS exam_date,
   datediff(r301.event_date, r101.appln_filing_date)
   AS days_to_exam
FROM
   wind_patents wp
      INNER JOIN
   reg101_appln r101 ON wp.appln_id = r101.appln_id
      INNER JOIN
   reg301_event_data r301 ON r101.id = r301.id
WHERE
   event_code = '0009185'
ORDER BY
   days_to_exam ASC;
```

Table 4 reports the first five results of Query 4. Application ID 8005567 was filed on March 26[th], 2008 and the applicant paid the fees to request examination 233 days later, on November 14[th], 2008.

**Table 4. First five records of output of Query 4**

| id | appln_id | appln_filing_date | exam_date | days_to_exam |
|---|---|---|---|---|
| 8005567 | 189424 | 2008-03-26 | 2008-11-14 | 233 |
| 9169460 | 267229289 | 2009-09-04 | 2010-05-28 | 266 |
| 9175409 | 273225234 | 2009-11-09 | 2010-08-13 | 277 |
| 3017319 | 16005751 | 2003-07-31 | 2004-05-07 | 281 |
| 8008488 | 55581546 | 2008-05-06 | 2009-03-13 | 311 |

*2.5 Attorney information*

PATSTAT Register also contains information on legal representatives (also called patent attorneys or patent agents). Such data have seldom been used but offer great potential for research. Recent uses include Somaya, Williamson and Zhang (2007), Wagner, Hoisl and Thoma (2014), and de Rassenfosse and Raiteri (2016). For example, Wagner, Hoisl and Thoma (2014) provide empirical evidence that external patent attorneys help facilitate the acquisition of distant knowledge.

Formally, applicants that have a residence or a place of business in an EPC member state do not have to rely on a legal representative. In practice, we have estimated that less than 10 per cent of EP applications do not have a representative. Note that a representative may be external to the firm (patent attorney firm) or internal (legal department of the applicant's company).

Query 5 identifies the oldest bulletin for each ID and the agent listed in that bulletin; hence, we extract information on the agent that has filed the application. The SELECT statement in parenthesis in Query 5 uses table REG102_PAT_PUBLN to identify the earliest publication—normally associated with a publication kind A1 or A2—in the bulletin. It then creates an identifier for the corresponding bulletin (*e.g.*, 200130 for the bulletin of week 30 of the year 2001). The information is stored in table P. Next, the query relies on table REG107_PARTIES to identify the agent information (TYPE = 'R') listed in the bulletin that matches the bulletin identifier.

**Query 5**
```sql
SELECT
   r107.id, p.appln_id, r107.bulletin_year, r107.bulletin_nr,
   r107.name
FROM
   reg107_parties r107
      INNER JOIN
   (SELECT r102.id, r101.appln_id,
         min(concat(cast(bulletin_year AS CHAR),
                   cast(bulletin_nr AS CHAR)))
         AS bulletin_first_publication
   FROM
      reg102_pat_publn r102
         INNER JOIN
      reg101_appln r101 ON r101.id = r102.id
         INNER JOIN
      wind_patents wp ON wp.appln_id = r101.appln_id
   GROUP BY
      r102.id, r101.appln_id) p
   ON r107.id = p.id
      AND concat(cast(r107.bulletin_year AS CHAR),
                cast(r107.bulletin_nr AS CHAR))
         = p.bulletin_first_publication
WHERE
   type = 'R'
ORDER BY
   r107.id;
```

As table 5 suggests, application ID 100008 was associated with patent attorney firm Strehl Schübel-Hopf & Partner as first legal representative. This information was published in the official bulletin on week 30 of year 2000.

**Table 5. First five records of output of Query 5**

| id | appln_id | bulletin_year | bulletin_nr | name |
|---:|---:|---:|---:|---:|
| 100008 | 15706408 | 2000 | 30 | Strehl Schübel-Hopf & Partner |
| 101623 | 15707784 | 2000 | 31 | Helms, Joachim, Dipl.-Ing. Patentanwalt |
| 104763 | 15710451 | 2001 | 20 | Dr. Weitzel & Partner |
| 105677 | 15711201 | 2001 | 38 | Hilleringmann, Jochen, Dipl.-Ing., et al |
| 105759 | 15711268 | 2000 | 38 | Strehl Schübel-Hopf & Partner |

It is useful to note that there are two ways of tracking event dates in the PATENT Register. The first involves tracking the time at which the event was reported in the bulletin of the EPO. However, the time lag between an event and its publication in the bulletin is subject to some uncertainty. Besides, if the publication of an event in the bulletin is not mandatory (such as for a change of representative), then there will be no bulletin date available. The second way involves tracking the date when the information has been added or changed in the respective IT system of the EPO (CHANGE_DATE attribute used in various tables). This date is closer to the actual occurrence of the event and easier to handle than the combined attributes BULLETIN_YEAR and BULLETIN_DATE.

*2.6 Searching for legal events: oppositions, revocations and limitations*

Legal events challenging the validity of patents provide information about patent value (Harhoff, Scherer and Vopel 2003, van Zeebroeck 2011) and market dynamics (Harhoff and Reitzig 2004). The PATSTAT Register contains information on three legal events that may affect the scope or validity of a patent: opposition, request for revocation and request for limitation. According to Article 99 of the EPC, any member of the public may file an opposition within nine months from when the mention of the grant of the European patent is published in the European Patent Bulletin. According to EPC Article 105a, only the owner may file a request for limitation or revocation. Such requests are very rare events.

Query 6 counts the occurrence of such events for applications in our sample. As the query indicates, there are eight values of the attribute EVENT_CODE from table REG301_EVENT_DATA that are relevant in this context.[7] Note that some EVENT_CODES have the same textual description (*e.g.*, EPIDOSCRVR1 and EPIDOSCRVR6 are associated with text "Change: proprietor files request for revocation"), although these codes have different internal meanings for the EPO. The description of codes is available in table REG402_EVENT_TEXT.

**Query 6**
```
SELECT
   r101.id, r101.appln_id, count(r301.id) AS nb_events
FROM
   wind_patents wp
      INNER JOIN
   reg101_appln r101 ON wp.appln_id = r101.appln_id
      INNER JOIN
   reg301_event_data r301 ON r101.id = r301.id
WHERE
   r301.event_code IN('0008299OPPO','0009260','EPIDOSCLIM1',
   'EPIDOSCRVR1','EPIDOSCRVR6','EPIDOSNLIM1','EPIDOSNRVR1',
   'EPIDOSNRVR6')
GROUP BY
   r101.id, r101.appln_id
ORDER BY
   nb_events DESC;
```

Table 6 shows that application ID 3711857 is associated with three events, all being oppositions filed in August/September 2013 (not reported).

---

[7] There are about 80,000 applications associated with at least one of these codes in the whole REG301_EVENT_DATA table.

**Table 6. First five records of output of Query 6**

| id | appln_id | nb_events |
|---:|---:|---:|
| 3711857 | 16039187 | 3 |
| 954452 | 15780379 | 3 |
| 1927846 | 15863176 | 3 |
| 7021392 | 57535 | 3 |
| 7723881 | 55002153 | 2 |

Users may also be interested in the outcome of these legal events. Outcomes of these three events are legal events themselves and are therefore recorded in the same table (REG301_EVENT_DATA). Relevant information is also scattered in separate tables, such as REG101_APPLN, REG130_OPPONENT, REG128_LIMITATION, REG114_DATES, REG201_PROC_STEP, REG203_PROC_STEP_DATE and TLS221_INPADOC_PRS (TLS231_INPADOC_LEGAL_EVENT in later versions of PATSTAT). For example, if the outcome of an opposition is a limitation, and not a complete revocation, then there will be a B2 publication in table REG101_APPLN. If a request for limitation has been granted, then there will be a B3 publication.

*2.7 Number of EPC member states in which the patent was validated*

In broad terms, a patent family refers to a group of patent applications that are all related to each other by way of one or several common priority filings (Martínez 2011 provides more precise definitions). The patent family size is often used as proxy for patent value (Putnam 1996, Harhoff, Scherer and Vopel 2003). One measure of family size that can be of interest to economic and management scholars is the so-called "geographic family size". It is obtained by counting the number of jurisdictions identified in a family. de Rassenfosse, Dernis and Boedt (2014) explain how to count the geographic family size for patents filed in different patent offices using PATSTAT. Often, however, researchers are interested in counting the geographic family size for patents filed at the EPO only. Indeed, once a patent is filed at the EPO, its family size can range from 1 to 38, where 38 is the number of EPC member states as of Autumn 2016.

At the time of filing, applicants must indicate the EPC member states where patent protection is sought with the payment of designation fees. The PATSTAT Register contains information on designated states. Information is straightforward to recover from table REG109_DESIGN_STATES. However, merely designating a country does not indicate that the patent entered into force in that jurisdiction. For that purpose, applicants must pay so-called validation fees. PATSTAT Register does not include information on the payment of validation fees; instead, it is necessary to look at the INPADOC data (table TLS221_INPADOC_PRS or TLS231_INPADOC_LEGAL_EVENT).

Query 7 provides an example of how to count the geographic family size for EP applications. The idea is to exploit information on the payment of annual fees to infer that a patent was validated in a jurisdiction. In particular, attribute PRS_CODE takes value "PGFP" indicating that the annual post-grant fee was paid for the country listed in field L501EP. Normally only the last payment is recorded. But because of data errors, the query uses the DISTINCT clause in the COUNT() function to avoid double counting for these erroneous cases. Information on renewal data is available in attribute L520EP, which indicates the year for which the annual payment has been paid (not exploited in this query).

```sql
                                                                              Query 7
SELECT
   r101.id, r101.appln_id,
   count(DISTINCT l501ep) AS nb_validated_states
FROM
   reg101_appln r101
      INNER JOIN
   wind_patents wp ON r101.appln_id = wp.appln_id
      INNER JOIN
   tls221_inpadoc_prs t221 ON r101.appln_id = t221.appln_id
WHERE
   prs_code = 'PGFP'
GROUP BY
   r101.id, r101.appln_id
ORDER BY
   nb_validated_states DESC, r101.id ASC;
```

Table 7 reports the first five records of Query 7. It shows that the maximum number of EPC member states in which a patent from the sample was validated is 33 (corresponding to ID 8001625).

**Table 7. First five records of output of Query 7**

| id | appln_id | nb_validated_states |
|---|---|---|
| 8001625 | 16417372 | 33 |
| 9007526 | 57755316 | 31 |
| 10710316 | 283793618 | 31 |
| 10720764 | 315926506 | 31 |
| 7728985 | 55301576 | 30 |

*2.8 Patent applicant's familiarity with EPO procedure*

PATSTAT Register data can also be used to generate truly novel indicators. Patent applicants widely differ in their use of the patent system as well as in their ability to deal with the patenting process. Some applicants may be highly sophisticated and take every procedural step they can to push their applications through the patent system or prevent the granting of patents to others. Table REG201_PROC_STEP stores a large number of interactions between the EPO and the applicants. Table REG202_PROC_STEP_TEXT provides a textual description of the codes used in REG201_PROC_STEP. It is clear that not all procedural steps reflect applicant "proficiency". The same step may reveal a strategic behaviour from the applicant or, on the contrary, a lack of professionalism. For instance, the payment of a surcharge for late payment may signal that the applicant consciously delayed examination or, on the contrary, that the applicant simply forgot to pay the fees.

The SELECT statement in Query 8 counts the total number of procedural steps (from table REG201_PROC_STEP) associated with the patent applications of the focal applicants and divides this number with the number of patent applications associated with the focal applicants. Focal applicants are the ones listed as the latest applicants in REG107_PARTIES (attribute IS_LATEST) for patent applications in WIND_PATENTS table.

**Query 8**
```sql
SELECT
   DISTINCT r107.name,
   round(count(*) / count(DISTINCT r201.id), 2) AS avg_proc_steps
FROM
   wind_patents wp
      INNER JOIN
   reg101_appln r101 ON wp.appln_id = r101.appln_id
      INNER JOIN
   reg201_proc_step r201 ON r101.id = r201.id
      INNER JOIN
   reg107_parties r107 ON r107.id = r201.id
WHERE
   r107.type = 'A'
      AND r107.is_latest = 'Y'
GROUP BY
   r107.name
ORDER BY
   avg_proc_steps DESC,
   r107.name;
```

The results are presented in Table 8. Applicants shown in the first five records all have only one patent application in the sample of wind patents (not reported). The patent by Neuhäuser GmbH went through 57 procedural steps. A closer look at the data (not reported) suggests that many of these steps involved STEP_CODE "PFEE" and "LOPR", which correspond to the payment of penalty fees (*e.g.*, surcharge for late payment) and the loss of particular rights.

**Table 8. First five records of output of Query 8**

| name | avg_proc_steps |
|---|---|
| Neuhäuser GmbH | 57.00 |
| HAWE Hydraulik GmbH & Co. KG | 56.00 |
| Panasonic Corporation | 56.00 |
| Sime-Stromag | 54.00 |
| Scialli, Aniello | 53.00 |

## 3. Concluding thoughts

The genesis for the present article is the popularity of the paper presenting an "Introduction to the PATSTAT database" (de Rassenfosse, Dernis and Boedt 2014). The paper was very well received by the community and has contributed to training a large number of young scholars and practitioners in the field of economics and management of innovation. We hope that the present introduction to the PATSTAT Register will similarly spur a community of practice that will advance knowledge in the field.

Although PATSTAT Register can be used on its own, its possibilities are best exploited when combined with the PATSTAT database. In general, PATSTAT Register is very attractive for researchers interested in particular procedural aspects or dynamics concerning ownership, inventorship and legal representation of EP publications. However, several perks of PATSTAT core are not available and/or not easy to add (*e.g.*, NUTS codes, sector information, harmonized names, *etc.*).

**Acknowledgements**

The authors are grateful to Ilona Ball, Laurie Ciaramella, Fabian Gäßler, T'Mir Julius, Albina Khairullina, Emilio Raiteri and Christian Soltmann for useful comments.

**References**

Arora, A., Gambardella, A., 1994. "The Changing Technology of Technical Change: General and Abstract Knowledge and the Division of Innovative Labour." *Research Policy* 23(5) 523–532.

Bösenberg, S., Egger, P., 2016. "R&D tax incentives and the emergence and trade of ideas." Paper presented at the 63rd Panel Meeting of Economic Policy hosted by the Nederlandsche Bank, April 2016.

Ciaramella, L., Martínez, C., Ménière, Y., 2016. "Tracking patent transfers in Europe: a first empirical analysis". Miméo, MINES ParisTech (Paris, France).

de Rassenfosse, G., Dernis, H., Boedt, G., 2014. "An introduction to the Patstat database with example queries." *Australian Economic Review* 47(3): 395–408.

de Rassenfosse, G., Palangkaraya, A., Webster, E., 2016. "Why do patents facilitate trade in technology? Testing the disclosure and appropriation effects." *Research Policy* 45(7): 1326–1336.

de Rassenfosse, G., Raiteri, E., 2017. "Technology Protectionism and the Patent System: Strategic Technologies in China".

de Rassenfosse, G., Zaby, A., 2016. The economics of patent backlog. Available at SSRN: http://dx.doi.org/10.2139/ssrn.2615090

Gans, J, Hsu, D., Stern, S., 2008. "The impact of uncertain intellectual property rights on the market for ideas: Evidence from patent grant delays." *Management Science* 54(5): 982–997.

Gäßler, F., Harhoff, D., 2016. "Patent Transfers in Europe – Data and Methodological Report". Miméo, Max Planck Institute for Innovation and Competition (Munich, Germany).

Harhoff, D., Reitzig, M., 2004. "Determinants of opposition against EPO patent grants—the case of biotechnology and pharmaceuticals." *International Journal of Industrial Organization* 22(4): 443–480.

Harhoff, D., Scherer, F., Vopel, K., 2003. "Citations, family size, opposition and the value of patent rights." *Research Policy* 32(8): 1343–1363.

Harhoff, D., Wagner, S., 2009. "The duration of patent examination at the European Patent Office." *Management Science* 55(12): 1969–1984.

Martínez, C., 2011. "Patent families: when do different definitions really matter?" *Scientometrics* 86(1): 39–63.

Mitra-Kahn, B., Marco, A., Carley, M., D'Agostino, P., Evans, P., Frey, C., Sultan, N., 2013. "Patent backlogs, inventories and pendency: An international framework." *UK Intellectual Property Office Working Draft* 2013/25.

Palangkaraya, A., Jensen, P., Webster, E., 2008. "Applicant behaviour in patent examination request lags." *Economics Letters* 101(3): 243–245.

Putnam, J., 1996. "The value of international patent rights", PhD thesis, Yale University.

Serrano, C. J., 2010. "The dynamics of the transfer and renewal of patents." *The RAND Journal of Economics* 41(4): 686–708.

Shane, S., 2002. "Selling university technology: patterns from MIT." *Management Science* 48(1): 122–137.

Shapiro, C., 1985. "Patent licensing and R&D rivalry." *The American Economic Review 75*(2): 25–30.

Somaya, D., Williamson, O., Zhang, X., 2007. "Combining patent law expertise with R&D for patenting performance." *Organization Science* 18(6): 922–937.

van Zeebroeck, N., 2011. "The puzzle of patent value indicators." *Economics of Innovation and New Technology* 20(1): 33–62.

van Zeebroeck, N., van Pottelsberghe de la Potterie, B., 2011. "Filing strategies and patent value." *Economics of innovation and new technology 20*(6): 539–561.

Wagner, S., Hoisl, K., Thoma, G., 2014. "Overcoming localization of knowledge—the role of professional service firms." *Strategic Management Journal* 35(11): 1671–1688.